\documentclass[aps,floats,twocolumn,prb,showpacs,floatfix]{revtex4-1}

\usepackage{latexsym,amsmath,amssymb,bm,graphicx,amsfonts,bbm}
\usepackage{graphicx,color}
\usepackage{subfigure}

\makeatletter\def\Dated@name{}\makeatother

\begin{document}

\title{Kinks in the periodic Anderson model}
\author{A. Kainz$^{1}$, A. Toschi$^1$, R. Peters$^2$ and K. Held$^1$}
\affiliation{$^1$ Institute for Solid State Physics, Vienna University of Technology,
1040 Vienna, Austria\\
$^2$ Department of Physics, Kyoto University, Kyoto 606-8502, Japan
}
\date{\today}

\begin{abstract}
Recently, dynamical mean field theory calculations have shown that \emph{kinks}  emerge in the real part of the self energy  of strongly correlated metals close to the Fermi level. This gives rise to a similar behavior in the quasi-particle dispersion relation as well as in the electronic specific heat.
Since f-electron systems are even more strongly correlated than the -hitherto studied- d-electron systems we apply the dynamical mean field 
approach   with the numerical renormalization group method  as impurity solver to study whether there are kinks in the periodic Anderson model.
\end{abstract}

\maketitle

In a general sense, kinks are an  abrupt but in reality often smooth crossover between two parameter regimes of a given physical system. 
Kinks are well known to result from the interaction between fermionic degrees of freedom and external bosonic modes:\cite{ashm} The coupling to collective excitations such as phonons influences the electronic dispersion. This results in kinks inside an energy range of the order of the Debye frequency $\omega_D$ centered around the Fermi level $E_F$. Typically these kinks are found at $\pm$40-70meV.\cite{lowenergykinks}

Due to the restriction to this rather small energy-range determined by $\omega_D$, the coupling to phonons cannot be the source of the observed high-energy kinks in the dispersion at energies $>80$meV.  Such high-energy kinks have been found, among others, in cuprates.\cite{highenergykink} In these cases, a mechanism not depending on interactions with external bosonic degrees of freedom has to be the microscopic origin.

It was recently discovered \cite{Nekrasov05a,kinkdisp} that such high-energy kinks emerge as an intrinsic feature of strongly correlated metals, in the real part of the self energy. The mechanism \cite{kinkdisp} indeed requires no additional coupling to external collective excitations and the corresponding kinks can arise at energies as high as a few hundred meV.

A mathematical understanding\cite{kinkdisp} can be gained on the basis of the  Hubbard model within  dynamical mean-field theory  (DMFT).\cite{Metzner89a,Georges92a,dmftrev} In the correlated metallic regime with a characteristic three peak spectral function, the kink energies can be shown to be dependent only on the renormalization strength $Z_{FL}$ and the non-interacting bandstructure.\cite{kinkdisp} For a Bethe-lattice with bandwidth $W=2D$ the kinks are located at
\begin{equation}
  \omega^\star_\pm= Z_{FL}(\sqrt{2}-1)D. \label{kinkloc2}
\end{equation}
The effective dispersion before and after this kink follows directly by a renormalization of the free dispersion $\varepsilon_{\mathbf k}$ and is given by
\begin{equation}
  E_{\mathbf k} = \begin{cases}
    c_-+Z^-_{CP} (\varepsilon_{\mathbf k}-\mu_0)   &    \text{for} \quad \omega<-\omega^\star_- \\
    Z_{FL} (\varepsilon_{\mathbf k}-\mu_0)   &    \text{for} \quad -\omega^\star_-<\omega<\omega^\star_+ \\
    c_++Z^+_{CP} (\varepsilon_{\mathbf k}-\mu_0)   &    \text{for} \quad \omega> \omega^\star_+ \\
  \end{cases} \; . \label{disp2}
\end{equation}
Here, $\mu_0$ denotes the chemical potential in the non-interacting case
and $Z_{CP}$ is a second, weaker renormalization factor for quasiparticle energies beyond $\omega^\star$.

The kinks in the real part of the self energy were also shown to reflect in a maximum in the susceptibility,\cite{Uhrig} and to result in corresponding kinks in the low temperature electronic specific heat $C_V(T)$.\cite{kinkcv} For the  latter, one can estimate a kink temperature
\begin{equation}
  T^\star \approx \frac{1}{5}(\sqrt{2}-1)Z_{FL}D. \label{tstar}
\end{equation}
    
Unfortunately, this kink temperature is for transition metals usually very 
large, i.e. of ${\cal O}(1000K)$. At such high temperatures the specific heat is dominated  by its phonon contribution, making an analysis virtually impossible.
However, there is the important exception of LiV$_2$O$_4$, the first heavy Fermion system with d-electrons,\cite{LiVO}  and a kink temperature of about 10K, confirming the theory.\cite{kinkcv} Naturally, one would hence look at f-electron systems 
with a similarly low energy scale.
However, at present it is unclear whether the
electronic kinks of Ref.\ \onlinecite{kinkdisp} are to be expected at all for f-electron
systems. They do  exist  for the Hubbard model \cite{kinkdisp} but not
for a single impurity Anderson model with a constant conduction electron density of states.\cite{Hewson} Hence we ask ourselves:
Are there purely electronic kinks in the periodic Anderson model,
the most fundamental model for f-electron systems?


This paper addresses directly this question and shows the emergence of kinks in the real part 
of the self energy Re$\Sigma(\omega)$ of the periodic Anderson model
 and the resulting effective energy-momentum dispersion $E_{\mathbf k}$.
The outline is as follows:
In Section \ref{mm},  we first introduce the periodic Anderson model used for the  analysis as well as the DMFT calculations  themselves, employing the numerical renormalization group (NRG) as an impurity solver.\cite{NRGref} In Section \ref{results} the results for the self energy $\Sigma(\omega)$, dispersion $E_{\mathbf k}$,  and  specific heat $C_V(T)$
are discussed.  The main results are summarized in Section \ref{Sec:conclusion}.


\section{Model and Methods} \label{pam}\label{mm}

\begin{figure}[t]
  \includegraphics{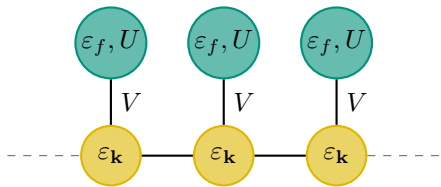}
  \caption{(Color Online) Depiction of the PAM. The interacting f-level ($\varepsilon_f,U$) of each site couples via the hybridization $V$ to the conduction band $\varepsilon_{\mathbf k}$. Direct hopping between different f-levels is not possible.}
  \label{fig:pam}
\end{figure}

The focus of this work lies on systems of strongly correlated f-electrons. These are confined to very narrow orbitals and interact with a local Coulomb repulsion $U$. Together with non-interacting conduction electrons these are the ingredients of the periodic Anderson model (PAM, Figure \ref{fig:pam}). The corresponding Hamiltonian  reads

\begin{align}
  H_{PAM} = &\sum_{\mathbf{k}\sigma} \varepsilon^{\phantom{+}}_\mathbf{k}a^+_{\mathbf{k}\sigma}a^{\phantom{+}}_{\mathbf{k}\sigma} +
  \varepsilon^{\phantom{+}}_f\sum_{i\sigma}f^+_{i\sigma}f^{\phantom{+}}_{i\sigma} \nonumber\\
  &+ \sum_{\mathbf{k}\sigma}V^{\phantom{+}}_{\mathbf{k}}(a^+_{\mathbf{k}\sigma}f^{\phantom{+}}_{\mathbf{k}\sigma} +
  f^+_{\mathbf{k}\sigma}a^{\phantom{+}}_{\mathbf{k}\sigma}) \label{andham} \\
  &+ U\sum_{i}f^+_{i\sigma}f^{\phantom{+}}_{i\sigma}f^+_{i\overline{\sigma}}f^{\phantom{+}}_{i\overline{\sigma}} -
  \mu\sum_{i\sigma}(f^+_{i\sigma}f^{\phantom{+}}_{i\sigma}+a^+_{i\sigma}a^{\phantom{+}}_{i\sigma}). \nonumber
\end{align}

Here, the operators $a,a^+$ represent a non-interacting conduction band with dispersion $\varepsilon_{\mathbf k}$ that can be thought of as a combination of s-, p- and d-bands, whereas the operators $f,f^+$ stand for the localized f-electrons with constant non-interacting energy $\varepsilon_f$. Each site of this model consists of a conduction band site as well as a f-orbital,  which are hybridized with each each other
by strength $V_{\mathbf k}$.  A detailed study of the PAM with DMFT(NRG) can be found, e.g., in Ref.\ \onlinecite{pam1999}. For the rest of this paper, the chemical potential $\mu$ is set to zero for the sake of simplicity.

As for the non-interacting case, the Hamiltonian Eq. (\ref{andham}) can be more conveniently written as a $2 \times 2$ orbital matrix for each $\mathbf{k}$-point (in terms of conduction and f-electron orbital). Due to the interaction $U$, the f-level is modified by a self energy $\Sigma(\omega)$, which is momentum-independent within DMFT. 
The corresponding one-particle Green's functions reads
\begin{equation}
  G_{\mathbf{k}}(\omega) = (\omega\mathbbm{1}-H_{\mathbf{k}})^{-1} =
  \begin{pmatrix}
    \omega-\varepsilon_f-\Sigma(\omega) & V_{\mathbf{k}} \\
    V_{\mathbf{k}}                             & \omega-\varepsilon_{\mathbf{k}}
  \end{pmatrix}^{-1}. \label{gdef1}
\end{equation}
For the calculations carried out in this work an energy $\varepsilon_c=\frac{1}{N_\mathbf{k}}\sum_{\mathbf k} \varepsilon_{\mathbf k}$ is considered in Eq.\ (\ref{andham}), \cite{pam1999} which can be interpreted as the center of mass of the conduction band and acts as an energy-shift of the conduction electrons. Additionally, for computational reasons, the momentum dependency of the hybridization is neglected for the rest of this work $V_{\mathbf k} \rightarrow V$. This leads to the following expressions for the local Green's functions
\begin{align}
  G_c(\omega) &= \int \frac{d^3k}{(2\pi)^3} \frac{1}{\zeta(\omega)-\varepsilon_{\mathbf k}} \label{glocc} \\
  G_f(\omega) &= \frac{1}{\omega-\varepsilon_f-\Sigma(\omega)} + \frac{V^2}{(\omega-\varepsilon_f-\Sigma(\omega))^2} G_c(\omega), \label{glocf} \\
  &\text{with} \,\, \zeta(\omega) = \omega-\frac{V^2}{\omega-\varepsilon_f-\Sigma(\omega)}. \label{zeta}
\end{align}
Eq.\ (\ref{glocf}) will serve as self-consistency relation for the DMFT calculations in this work.

\subsection*{DMFT self-consistency cycle} \label{scc}

The complex many-body problem is approximated by the DMFT \emph{self-consistency cycle}.\cite{dmftrev} For the PAM implementation, we need  to solve the integral in Eq.\ (\ref{glocc}). For the {\em Bethe lattice} 
with semi-elliptical density of states $N^{Bethe}(\varepsilon) = \frac{2}{\pi t^2} \sqrt{t^2-\varepsilon^2}$, the integral can be calculated analytically, yielding
\begin{align}
  G_c(\omega) = \frac{2 \zeta(\omega)}{t^2}\left(1-\sqrt{1-\frac{t^2}{\zeta(\omega)}}\right) \label{gloccbt} \; .
\end{align}

\begin{figure}[t!]
  \includegraphics{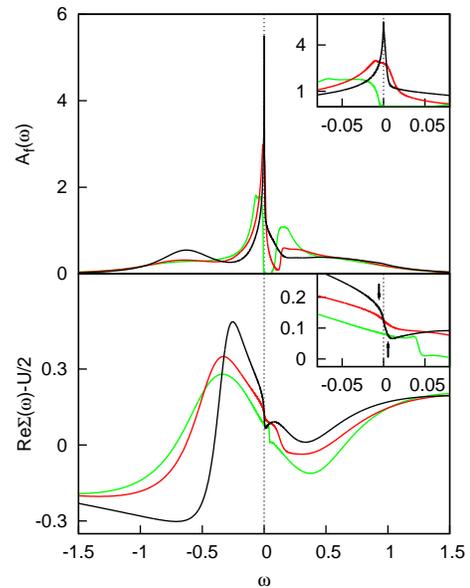}
  \caption{(Color Online) Upper panel: local f-electron spectral functions of the PAM on a simple cubic lattice for $\varepsilon_f=-0.5$, $U=1$, $V^2=0.1$, $W=2$ and different values of $\varepsilon_c=0.1,0.3,0.5$ (green/light gray, red/dark gray, black). Lower panel: corresponding real part of the self energy. Insets: zoom in around the Fermi level at $\omega=0$. The kinks are indicated by arrows for the case $\varepsilon_c=0.5$.}
  \label{fig:aec}
\end{figure}

\begin{figure}[t!]
  \includegraphics{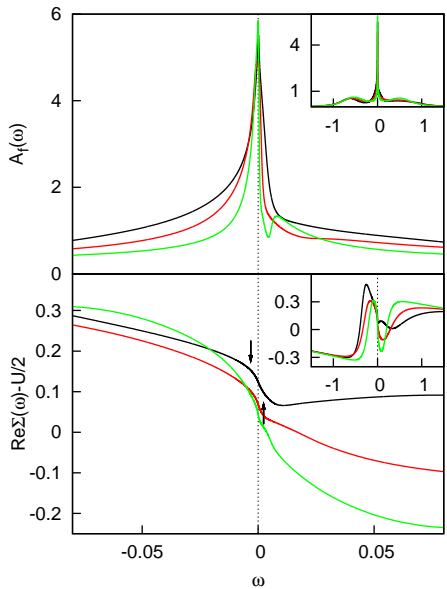}
  \caption{(Color Online) Upper panel: local f-electron spectral function 
for $U=-2\varepsilon_f=1$, $\varepsilon_c=0.5$, $V^2=0.1$ and different bandwidths $W=2,4,6$ (black, red/dark gray, green/light gray) for a simple cubic lattice.
Lower panel: corresponding  real part of the self energy. Insets: larger frequency window. The kinks are indicated by arrows for the case $W = 2$.}
  \label{fig:vglw}
\end{figure}

For  the {\em simple-cubic lattice}, integral Eq.\ (\ref{glocc}) can be rewritten in the form \cite{Joyce}
\begin{equation}
  G_c(\omega) = \frac{1}{\zeta(\omega)} P(z), \label{glocc2}
\end{equation}
with the abbreviation $z(\omega)=-\frac{W}{2\zeta(\omega)}$ ($W=12t$ is the bandwidth). Here, the function $P(z)$ is equivalent to a product of two hypergeometric functions of the type $_2F_1(1/2,1/2;1;k^2)$, namely
\begin{equation}
  P(z)=\frac{\sqrt{1-\frac{3}{4}x_1}}{1-x_1}\,{_2F_1}(\frac{1}{2},\frac{1}{2};1;k_+^2)\,{_2F_1}(\frac{1}{2},\frac{1}{2};1;k_-^2). \label{pz}
\end{equation}

The abbreviations $k_\pm,x_1$ and $x_2$ are defined by
\begin{align}
  k_\pm^2 &= \frac{1}{2} \pm \frac{1}{4}x_2\sqrt{4-x_2}-\frac{1}{4}(2-x_2)\sqrt{1-x_2} \\
  x_1 &= \frac{1}{2} + \frac{1}{6}z^2 - \frac{1}{2}\sqrt{1-z^2}\sqrt{1-\frac{1}{9}z^2} \\
  x_2 &= \frac{x_1}{x_1-1}.
\end{align}


Unfortunately, to treat the PAM with DMFT(NRG) involves overcoming numerical difficulties\cite{pam1999,NRGref} which affect the stability of the calculations (the NRG step in particular) as well as the convergence behavior of the DMFT-loop. Specifically, to arrive at fixed and stable solutions of the PAM for the considered parameter regime, it was necessary to introduce a small imaginary shift of the real frequencies,\cite{deltino} $\omega \to \omega + i\delta$, which takes care of possible delta peaks in the Green's function Eq.\ (\ref{glocf}) or hybridization function, and to make use of Broyden's method of convergence stabilization.\cite{broyd,broyd0}
The logarithmic NRG discretization has been taken to be $\Lambda=2$ with      checks for  $\Lambda=1.8, 1.9, 2.0$ yielding very similar results.

\section{Results}\label{results}
It is known that due to the hybridization of conduction and f-electrons, the PAM at half filling ($\varepsilon_f=-U/2$ and $\varepsilon_c=0$) represents a Kondo-insulator in the paramagnetic phase studied throughout this paper.\cite{pam1999} However, searching for kinks, a metallic configuration is necessary, and one with a well defined three peak spectral function desirable. These requirements can be achieved by breaking the particle-hole symmetry in such a way that the energy level of the conduction electrons is increased while the f-level is held \emph{symmetric},\cite{pam1999,NRGref} see  Fig. \ref{fig:aec} (upper panel).

\begin{figure}[t]
  \includegraphics[width=0.48\textwidth]{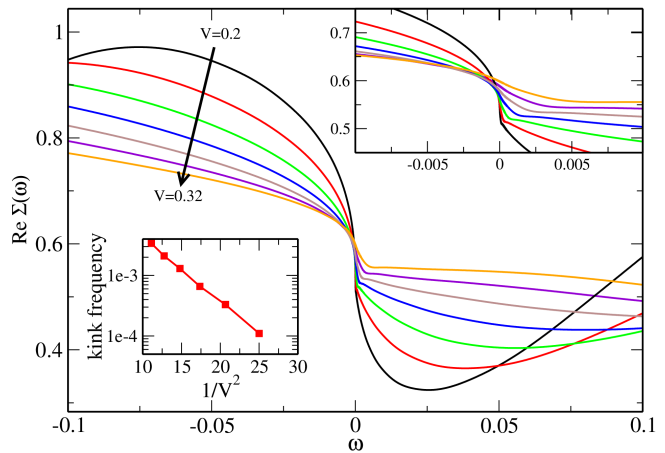}
\caption{(Color Online) Real part of the self energy 
for  the Bethe lattice with $U=-2\varepsilon_f=1$, $\varepsilon_c=0.5$, $W=2$ and various 
$V$. Right inset: magnification around the Fermi energy.
Left inset: frequency of the kink vs.\ inverse hybridization strength.}
  \label{fig:diff_V}
\end{figure}

\begin{figure*}[t!]
  \subfigure[] {\includegraphics[width=0.33\textwidth]{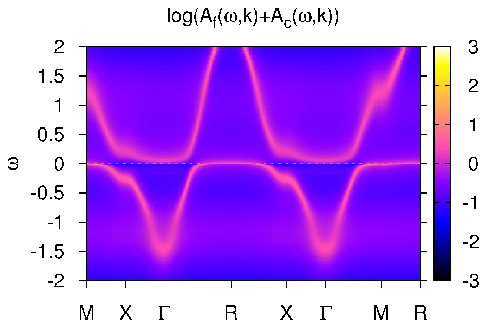}}
  \subfigure[] {\includegraphics[width=0.33\textwidth]{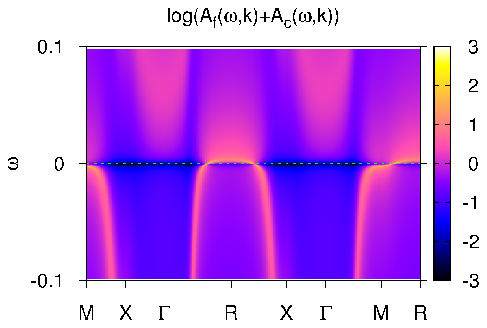}}
\subfigure[] {\includegraphics[width=0.33\textwidth]{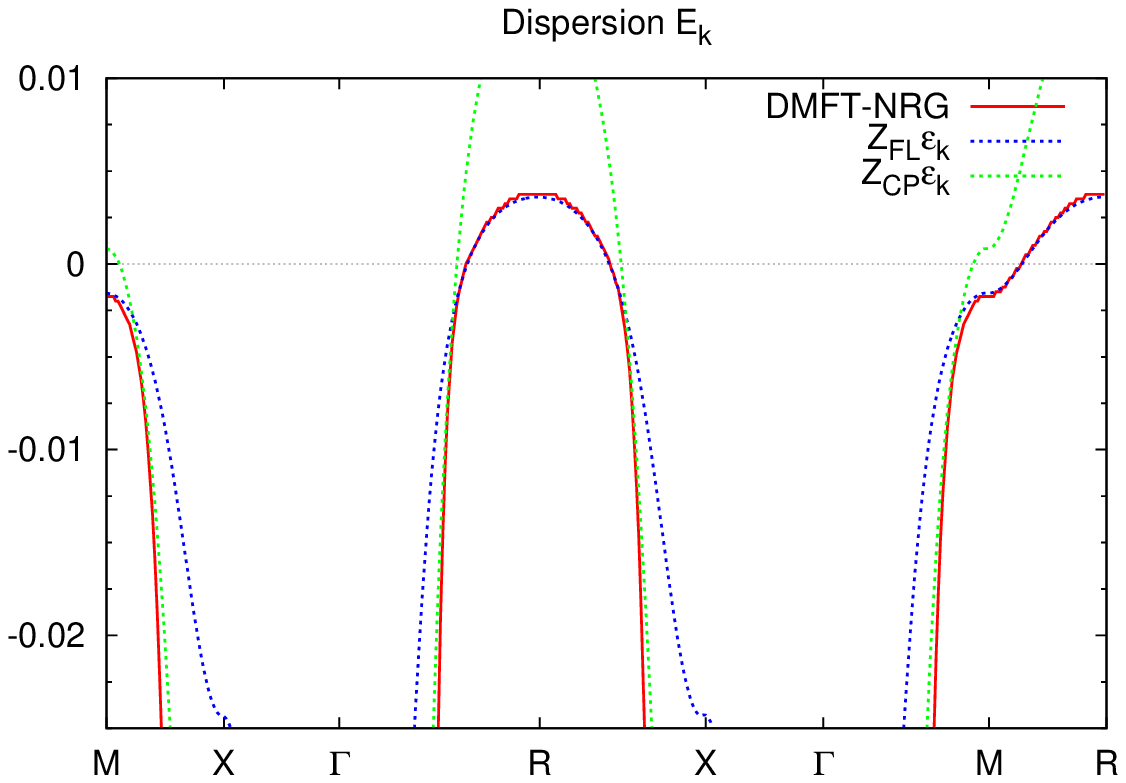}}
  \caption{(Color Online) (a) Overall spectral density $(A_f(\mathbf{k},\omega)+A_c(\mathbf{k},\omega))$ on a logarithmic scale, for  a simple cubic lattice with $U=-2\varepsilon_f=2$, $\varepsilon_c=0.5$, $V^2=0.2$, and $W=3.5$. (b) magnification around the Fermi energy. The curves along the maxima of (a) and (b) represent the dispersion $E_{\mathbf k}$. In (c) this dispersion (solid red curve) is compared to the non-interacting one multiplied with renormalization factors $Z_{CP}=0.202$ (green dotted) and $Z_{FL}=0.0689$ (blue dotted) extracted from a fit of the corresponding self energy. The dotted curves fit the DMFT(NRG) result very well, hence indicating kinks in the dispersion of the PAM.}
  \label{fig:disp1}
\end{figure*}

\begin{figure}[b!]
  \includegraphics[width=0.33\textwidth]{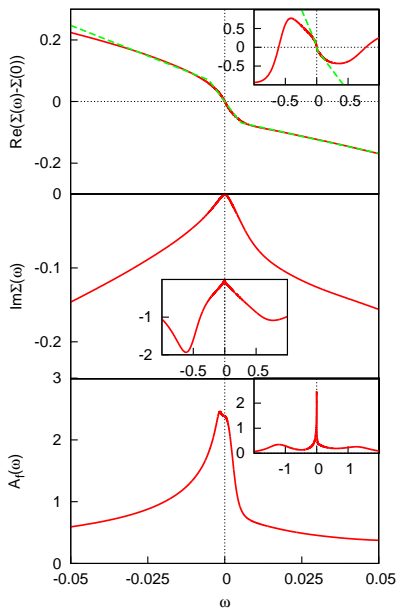}
  \caption{(Color Online) Upper panel:  real part of the self energy for the parameters   $U=-2\varepsilon_f=2$, $\varepsilon_c=0.5$, $V^2=0.2$, and $W=3.5$, together with  a piecewise linear fit (green dashed  line).
Middle panel: corresponding imaginary part of the self energy. Lower panel: corresponding local spectral function. 
}
  \label{fig:res2}
\end{figure}

The effect of the conduction band shift $\varepsilon_c$ on the real part of the self energy is illustrated in the bottom panel of Fig. \ref{fig:aec}.
For $\omega<0$, the overall behavior of the self energy is the same  as one would expect for a strongly correlated metal. The real part shows a basically
linear behavior for small energies and eventually reaches its maximum. After that, it falls off rapidly and ultimately converges to a constant.

For $\omega>0$, on the other hand, $\Sigma(\omega)$ experiences the consequences of the hybridization.
 The hybridization gap, which moves to higher frequencies for increasing $\varepsilon_c$, is reflected in the imaginary part of the self energy as a second  minimum
 of  $|$Im$\Sigma(\omega)|$ (besides $\omega=0$; not shown). 
In the Kramers-Kronig related  Re$\Sigma(\omega)$ it shows up as an inflection point in
Fig.\ \ref{fig:aec}.

Compared to these coarse features, kinks are fine structures which have been
overlooked in the Hubbard model prior to Refs.\  \onlinecite{Nekrasov05a} and 
  \onlinecite{kinkdisp}, and
in the periodic Anderson model up to this point.
A closer inspection of the self-energy shows a kink for  $\omega<0$  close to the Fermi energy, see the inset of
Fig.\ \ref{fig:aec}, particularly well visible for  $\varepsilon_c=0.5$.
Upon increasing the bandwidth $W$ and hence decreasing the density of states of the conduction electrons at the Fermi level, the width of the central Abrikosov-Suhl resonance is reduced, as is the kink energy, cf.\ $W=4$ in Fig.\ \ref{fig:vglw}.

For  $\omega>0$, the inflection point (hybridization gap) 
makes the identification of a kink   more complicated. The most clear separation of
kink (at $\omega \sim 0.004$) and inflection point (at $\omega\sim 0.01$) is arguably
obtained for   $W=4$
 in Fig.\ \ref{fig:vglw}. But also for $W=6$ two distinct features are 
discernible for $\omega>0$.


Keeping the bandwidth fixed and modifying instead the hybridization strength, we
show in Fig.\ \ref{fig:diff_V} the real part of the self energy for the Bethe lattice. A kink for  $\omega>0$ is well visible, in particular for a smaller
hybridization $V$, see right inset of  Fig.\ \ref{fig:diff_V}.
In the left inset of  Fig.\ \ref{fig:diff_V} we plot the  kink frequency of the upper panel vs.\ the inverse hybridization strength $1/V^2$ on a logarithmic scale. This reveals that the kink frequency shows the same exponential dependence on $1/V^2$ as the Kondo temperature.

Let us note that compared to the Hubbard model  for d-electron systems the kinks of the PAM are located at much lower energies (by about one order of magnitude). This stems from the very small quasiparticle renormalization $Z_{FL} \ll 1$, which is not surprising, since the interaction strength is more enhanced in f-systems due to the confinement of electrons in the very narrow f-orbitals.

\subsection*{Dispersion relation} \label{disp}

After finding kinks in the real part of the PAM self energy, let us now  investigate if they have a similar influence on the effective dispersion relation $E_\mathbf{k}$ as for the Hubbard model. Here,  $E_\mathbf{k}$ is  defined as the maximum of the $k$-resolved spectrum  $A(\mathbf{k},\omega)$ with respect to  $\mathbf{k}$, as in angular-resolved photo-emission experiments. Since the PAM is an effective two band model, it has two such dispersion relations for f and c(onduction) electrons.
 In Fig. \ref{fig:disp1} (a) and (b) the overall spectral density $A_f(\mathbf{k},\omega)+A_c(\mathbf{k},\omega)$ is plotted. The dispersion extracted from these spectral functions is depicted in Fig. \ref{fig:disp1} (c) (red curve)
At first glance, no kink feature is discernible.

On the other hand, the kink in the real part of the self energy
should directly reflect in a kink of the dispersion relation, whereas the imaginary part smears out the maxima.
This can be demonstrated by employing a linear fit to the real part of
 the self energy in  Fig. \ref{fig:res2}.
Taking into account the frequency range $\omega \in [-0.05,0.05]$, we obtain
a Fermi liquid renormalization factor $Z_{FL}=0.0689$ for the slope around
the Fermi level,
 and two  renormalization factors  $Z^{+}_{CP}=0.202$ and $Z^{-}_{CP}=0.312$ for the slopes after the kink at  $\omega^\star_-\approx-0.005$ and $\omega^\star_+\approx 0.005$, respectively; cf.\  Eq.\ (\ref{disp2}).
These renormalization factors are related to the corresponding 
self energy slopes $\partial \Sigma/\partial \omega$ as $Z=(1-\partial \Sigma/\partial \omega)^{-1}$. 

From the Fermi liquid renormalization $Z_{FL}$ and from  $Z^{-}_{CP}$ the
two dashed dispersion relations  in Fig.\  \ref{fig:disp1} (c) are derived.
These describe the NRG dispersion relations accurately around the Fermi level and for more negative frequencies. In-between there is  a crossover 
from one curve to the other. This  reflects the self-energy kink, which due to the
already strong curvature of the non-interacting (or renormalized) 
dispersion does however not show up as an abrupt change of slope.


\subsection*{Specific heat} \label{cv}

\begin{figure}[t!]
  \includegraphics[width=0.47\textwidth]{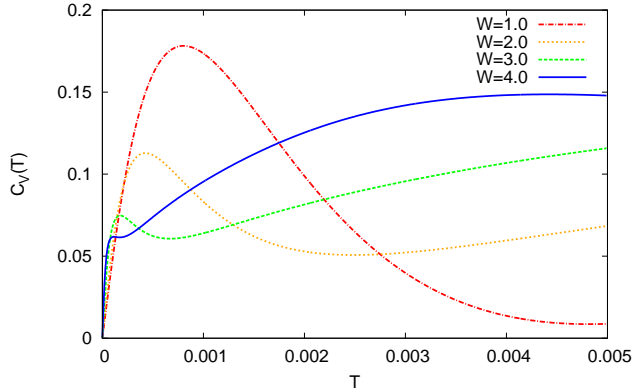}
  \caption{(Color Online) Specific heat of a  Bethe-lattice  PAM with
$U=-2\varepsilon_f=1$, $\varepsilon_c=0.5$, $V^2=0.1$ and various bandwidths.
}
  \label{Fig:Cv}
\end{figure}
The kinks in the self energy can also be expected to reflect as a change of the linear behavior of the specific heat. As in Ref.\ \onlinecite{kinkcv},
we have employed the relation \cite{Abrikosov63} between low temperature entropy and spectral function, 
yielding the following conduction and f-electron contribution to  the specific heat at temperature $T$:
\begin{eqnarray}
C_V(T)&=&{2T}\int_{-\infty}^{\infty}{\rm d}y  \frac{y^2e^y}{(e^y+1)^2} \big[A_c(yT)+
\frac{A_f(yT)}{Z(yT)} \nonumber \\ && 
 + \frac{1}{\pi} {\rm Re} G_f(yT) {\rm Im} \Sigma(yT) \big] \; ,
\label{Eq:CV}
\end{eqnarray}
where $Z(yT)=(1-{\rm Re} \Sigma(yT))^{-1}$ is the renormalization factor for the $f$-electrons, and the summation over the two spin directions is accounted explicitly by the prefactor $2$. The last term (second line) in Eq.\ (\ref{Eq:CV})
also accounts for the imaginary part of the self energy and is beyond Ref.\ \onlinecite{kinkcv}.

Fig.\ \ref{Fig:Cv} shows the specific heat for the Bethe lattice and different bandwidths (and density of states) of the conduction electrons
calculated according to Eq.\  (\ref{Eq:CV}). At low temperatures, there is
a linear increase of the specific heat as is to be expected for a Fermi liquid. The next dominant feature is a minimum found, e.g., at $T\approx 0.0006$ for $W=3$. The origin of this minimum is the hybridization gap which leads to a reduced number of states in the corresponding energy interval.
These two dominant features hide a more delicate kink feature which
according to  Eq.\  (\ref{Eq:CV}) should be present (note $Z(yT)$  strongly depends on temperature), but is not discernible.

\section{Conclusion} \label{Sec:conclusion}
We have identified kinks in the real part of the self energy of the 
periodic Anderson model, the 
arguably simplest model for f-electrons systems.
The hybridization gap leads to an additional feature,  in our case at $\omega>0$, making the clear identification of the kink more difficult than for the Hubbard model. The kink frequency
follows the same exponential dependence on the hybridization strength as the Kondo temperature.
In contrast to the Hubbard model, the kink is difficult to detect visibly in the energy-momentum dispersion since the non-interacting energy-momentum dispersion
has a strong curvature around the kink energy. Similarly, also in the specific heat, the fingerprint of the self-energy kink is less obvious because a  stronger feature, associated with the hybridization gap, is superimposed.

\vskip 2mm

We thank A. Toth, K. Byczuk, M. Kollar and D. Vollhardt for discussions, and the Austrian Science Fund for financial support via Research Unit FOR 1346
(project ID  I597-N16, AK), project ID I610-N16 (AT), and the SFB ViCoM F4103-N13 (KH); as well as  the Japan Society for the Promotion of Science (JSPS)
for its support through the FIRST Program (RP).


\end{document}